\documentstyle[psfig]{mn}

\begin{document}
\title[Different cluster mass estimates]
{A comparison of different cluster mass estimates:
consistency or discrepancy ?}
\author[Wu, Chiueh, Fang and Xue]{Xiang-Ping Wu$^{1,2}$, 
                                 Tzihong Chiueh$^{1}$,
                                 Li-Zhi Fang$^{3}$ and
                                 Yan-Jie Xue$^{2}$ \\
$^1$Institute of Astronomy, National Central University, Chung-Li 
    32054, Taiwan\\
$^2$Beijing Astronomical Observatory, Chinese Academy of Sciences,
    Beijing 100012, China\\
$^3$Department of Physics, University of Arizona, Tucson, AZ 85721, U.S.A.
} 
\date{submitted 1998 April; accepted 1998 August}
\maketitle
\begin{abstract}

Rich and massive clusters of galaxies at intermediate redshift
are capable of magnifying and distorting the images of background galaxies.
A comparison of different mass estimators 
among these clusters can provide useful information about the
distribution and composition of cluster matter and their dynamical evolution. 
Using a hitherto largest sample of lensing clusters drawn from literature,
we compare the gravitating masses of clusters derived from  the
strong/weak gravitational lensing phenomena, from the X-ray measurements 
based on the assumption of hydrostatic equilibrium, and from the conventional 
isothermal sphere model for the dark matter profile characterized 
by the velocity dispersion and core radius of galaxy distributions in
clusters.
While there is an excellent agreement between the weak lensing, X-ray
and isothermal sphere model determined cluster masses, these methods 
are likely to underestimate the gravitating masses enclosed within 
the central cores of clusters by a factor of $2$--$4$ 
as compared with the strong lensing results.
Such a mass discrepancy has probably arisen from the inappropriate 
applications of the weak lensing technique and the hydrostatic 
equilibrium hypothesis to the central regions of clusters
as well as an unreasonably large core radius for both luminous 
and dark matter profiles. Nevertheless, it is pointed out that 
these cluster mass estimators may be safely applied on
scales greater than the core sizes. Namely,
the overall clusters of galaxies at intermediate redshift 
can still be regarded as the dynamically relaxed systems,
in which the velocity dispersion of galaxies and the temperature of
X-ray emitting gas are good indicators of the underlying gravitational 
potentials of clusters.

\end{abstract}


\vskip -3in

\section{Introduction}

Clusters of galaxies are the largest coherent and gravitationally
bound objects in the universe. They are often used for cosmological
test of theories of structure formation.
In particular, their matter composition, the mass-to-light
ratio $M/L$ (e.g. Bahcall, Lubin, \& Dorman 1995)
and the baryon fraction $f_b$ (e.g. White et al, 1993)
play a potentially important role
in the direct measurement of mean mass density of the
universe, $\Omega_m$. It appears that
the observational and theoretical results are likely to merge
nowadays, in favor of a low mass density universe of $\Omega_m\sim0.3$
(e.g.  Bahcall et al. 1995; Ostriker \& Steinhardt 1995). 
Yet, the key issue in such a ``direct'' measurement of
the cosmological density parameter is closely connected to
the question of how accurately one can determine the total gravitating
masses of galaxy clusters.
Recall that the traditional cluster estimators strongly rely upon the
assumption of hydrostatic equilibrium: both optical galaxies and
intracluster diffuse gas quasi-statically
trace the underlying gravitational
potential of the whole cluster. However, a number of recent X-ray observations
(e.g. Henry \& Briel 1995; Markevitch 1996; etc.) have argued that
the hydrostatic hypothesis is inapplicable to at least merging clusters.
Therefore, an independent method of estimating cluster mass
is highly appreciated. At present, it is widely recognized that 
the gravitational lensing is a unique and also ideal tool to fulfill 
the task, in the sense that it provides a fair estimate of cluster masses
regardless of the cluster matter components and their dynamical states.

In the framework of gravitational lensing,  mapping 
the gravitational potential of a galaxy cluster can be achieved by 
the detailed modeling of the strongly distorted and magnified arclike 
images inside the core of the cluster (Fort \& Mellier 1994), 
by the statistical analysis of the weakly distorted images of faint and 
distant galaxies around the cluster (Kaiser \& Squires 1993), and  
by counting the populations of distant galaxies 
(Wu \& Hammer 1995; Taylor et al. 1998) and quasars (Wu \& Fang 1996a;
references therein) around the cluster. A combination of these lensing
phenomena would, in principle, allow one to depict  an overall matter 
distribution of cluster on scales spanning two decades
from the inner core of $\sim 30$ kpc to the outer regime
of $\sim 10$Mpc (Wu \& Fang 1996b).  However, in the practical 
measurements the lensing signals often 
appear to be relatively weak, and a precise determination of cluster masses
suffers from the limitations of instrument sensitivities and viewing fields.
In particular, for different lensing phenomena
different techniques have been developed and employed in 
the reconstruction of matter distributions of clusters.
Hence,  before one proceeds to the cosmological applications of 
the gravitating masses of clusters derived from gravitational lensing,  
it is also worthy of examining whether different approaches based on
different lensing phenomena yield a consistent gravitating  mass of cluster.

On the other hand, apart from the doubt as to 
whether the gravitational lensing provides an accurate
cluster mass estimate, it is of great interest to compare 
the cluster masses derived from 
traditional method with those from gravitational lensing.
Previous comparison has been made among many individual clusters, 
in which both strongly/weakly distorted images of background galaxies
and X-ray/optical emissions are detected.  
Basically, a good agreement  between the  
X-ray and weak lensing determined cluster masses 
has been detected (Squires et al. 1996, 1997b; Smail et al. 1997;
Allen 1998), in contrast to the situation of strong lensing, in which
the lensing derived cluster masses are typically 
 $\sim2-4$ times larger than the cluster dynamical masses obtained
under the assumption of hydrostatic equilibrium
(e.g. Wu 1994; Miralda-Escud\'e \& Babul 1995; 
Allen, Fabian, \& Kneib 1996; Markevitch 1997).
This argument is further strengthened by Wu \& Fang (1997) using
a large sample of 29 lensing clusters drawn from literature.
Several attempts have correspondingly been made to explain
the discrepancy such as the possible contributions of nonthermal
pressure (Loeb \& Mao 1994; Ensslin et al. 1997) and projection effect 
(Miralda-Escud\'e \& Babul 1995; Girardi et al. 1997b).  
An important progress in 
understanding the issue has been made recently by Allen (1998), based on
a detailed comparison of the cluster mass determinations  
from X-ray and gravitational lensing for 13 lensing clusters. 
He showed that the mass
discrepancy between the X-ray and strong lensing analyses exists only
in the non-cooling flow clusters, which can be well accounted for by 
the significant offsets between the X-ray and strong lensing centers.
This implies that in the central regions of these clusters, 
the traditional ``quasi-static'' hypothesis for the X-ray  emitting gas 
has probably broken down.  However, it should be noticed that even for these 
non-cooling clusters the weak lensing determined cluster masses agree
nicely with the X-ray determined ones.

The current comparisons of different cluster mass estimates 
seem to suggest that the reported mass discrepancy 
between X-ray and gravitational lensing analyses may, at least partially, 
arise from the employments of different lensing techniques: 
Strong lensing and weak lensing yield an inconsistent cluster mass.
Although  this disagreement could be attributed to observations because  
strong lensing and weak lensing measurements probe the different 
regions of clusters, a direct comparison between cluster masses 
obtained from these two lensing phenomena has now become possible
with the progress of weak lensing technique and the increasing
population of arcs/arclets. If the cluster masses derived from 
strong lensing do systematically exceed the weak lensing values, 
one would be faced with the following
difficulties: The consistency of the weak lensing and X-ray determined 
cluster masses indicates that the strong lensing analysis may lead to 
an overestimate of gravitating  mass for cluster, which requires 
a re-examination of the modeling of arcs/arclets as the
cluster mass estimator. On the other hand, the strong lensing phenomena 
(arcs or arclets) are usually believed to be a good indicator of the 
underlying gravitational potential of the lensing clusters, 
in which there are almost no free parameters once the redshifts of
the arclike images are known. This leads us to the possibility that 
both weak lensing and X-ray measurements may underestimate
the true cluster mass by a factor of $2$--$4$. 
The latter point is directly related  to 
the question whether the previously estimated $M/L$ and $f_b$ of clusters 
gives rise to an underestimate of the cosmological density parameter 
$\Omega_m$.  Of cause, there may exist a third possibility that enables us
to reconcile the cluster mass difference while maintain
the validity of these different mass estimators. 
Alternatively, an examination of the consistency or 
discrepancy between the strong/weak lensing and X-ray/optical mass measurements 
of clusters will be of great help to our understanding of dynamical and 
evolutionary properties of rich clusters at intermediate redshifts.
Throughout this paper, we assume $H_0=50$ km s$^{-1}$ Mpc$^{-1}$ and 
$\Omega_0=1$.

\section{Sample}

\begin{table*}
\vskip 0.2truein
\begin{center}
\caption{Strong Lensing Cluster Sample}
\vskip 0.2truein
\begin{tabular}{ l l l l l l l }
\hline
cluster & $z_{cluster}$ & $T$ (keV) & $\sigma_{gal}$ (km s$^{-1}$) &
$z_{arc}$ & $r_{arc}$(Mpc) & $mass (10^{14}M_{\odot})$ \\
\hline
     &       &      &   &  &   &   \\
A370 & 0.373 & $7.1^{+1.0}_{-0.8}$ & $1367^{+310}_{-184}$ & 
                                1.3 & 0.35 & 13.0 \\
     &        &                      &                      &  
                                0.724 & 0.16 & 2.90 \\
A963 & 0.206 & $6.8^{+0.4}_{-0.5}$ & $1100^{+480}_{-210}$ &
                               ... & 0.0517 &0.25 \\
     &        &                      &                      &
                               0.711 & 0.080  & 0.60 \\
A1204 & 0.170 & $3.8^{+0.2}_{-0.2}$ & ... & ... & 0.0311 & 0.10\\
A1689 & 0.181 & $9.0^{+0.4}_{-0.3}$ & $1989^{+245}_{-245}$ &
                               ... & 0.183 & 3.6 \\
A1835 & 0.252 & $8.2^{+0.5}_{-0.5}$ &  & ...& 0.150 & 1.98\\
A1942 & 0.224 &  ...      &  ...    & ...& 0.0372 & 0.18 \\
A2104 & 0.155 &  ...      &  ...    & ...& 0.025  & 0.064 \\
A2163 & 0.203 & $13.9^{+0.7}_{-0.5}$ & 1680 & 0.728 & 0.0661 & 0.41 \\
A2218 & 0.171 & $7.0^{+1.0}_{-1.0}$ & $1405^{+163}_{-145}$ & 
                                      1.034 & 0.26 & 2.7 \\
      &        &           &        & 0.702 & 0.0794 & 0.623\\
      &        &           &        & 2.515 & 0.0848 & 0.570\\
A2219 & 0.228 & $11.8^{+1.3}_{-0.8}$ & ...  & ... &  0.079 & 0.517\\
      &        &                      &      & ... & 0.110 & 1.60 \\
A2280 & 0.326 &   ...    & $948^{+516}_{-285}$ & ... & 0.080 & 0.59\\
A2390 & 0.228 & $8.9^{+1.0}_{-0.8}$ & $1093^{+61}_{-61}$ & 
                                      0.913 & 0.177 & 2.54 \\
A2397 & 0.224 &   ...    &    ...    & ... &  0.0698 & 0.45\\
A2744 & 0.308  & $12.1^{+1.4}_{-1.0}$ & $1950^{+334}_{-334}$   &
                                        ... & 0.1196 & 1.136\\
A3408 & 0.0419 &  ...    &    ...    & 0.0728 & 0.0568 & 0.44 \\
S295  & 0.299 &    ...    & 907        & ... & 0.0329 & 0.14\\
      &        &           &            & 0.930 & 0.14   & 1.6\\
CL0024 & 0.391 & ...  & $1339^{+233}_{-233}$ & 1.390 & 0.214 & 3.324\\
CL0302 & 0.423 & ...  & 1100           & ... & 0.122  & 1.60 \\
CL0500 & 0.327 & $7.2^{+3.7}_{-1.8}$ & $1152^{+214}_{-214}$ & ... 
                                       & 0.15 & 1.90\\
CL2236 & 0.552 & ...    &    ...       & 1.116 & 0.0876 & 0.30 \\
CL2244 & 0.328 & $6.5^{+1.8}_{-1.3}$    &    ...       
                                      & 2.236 & 0.0465 & 0.20 \\
MS0440 & 0.197 & $5.3^{+1.3}_{-0.8}$ & $606^{+62}_{-62}$ & 
                                      0.530 & 0.089 & 0.89\\
MS0451 & 0.539 & $10.2^{+1.5}_{-1.3}$ & $1371^{+105}_{-105}$& 
                                      ... & 0.190 & 5.2\\
MS0955 & 0.145  &  ...   &     ...   & ... & 0.0385 & 0.22 \\
MS1006 & 0.261 &  ...   & $906^{+101}_{-101}$ & ... & 0.079 & 0.57\\
       &        &     &                     & ... & 0.14  & 1.8 \\
       &        &     &                     & ... & 0.28  & 7.2 \\
MS1008 & 0.306 & $7.3^{+2.5}_{-1.5}$& $1054^{+107}_{-107}$& 
                                             ... & 0.26 & 6.1 \\
MS1137 & 0.783 & ...  &   ...          &  ... & 0.044 & 0.19 \\
       &        &     &              &  ... & 0.147 & 2.1 \\
       &        &     &              &  ... & 0.151 & 2.2 \\
MS1358 & 0.329 & $6.5^{+0.7}_{-0.6}$ & $937^{+54}_{-54}$ & 
                                        4.92 & 0.121 & 0.827\\
MS1445 & 0.257 & $5.6^{+0.2}_{-0.3}$ & $1133^{+140}_{-140}$& 
                                        ... & 0.098 &0.86 \\
MS1621 & 0.427 & ...   & $793^{+55}_{-55}$ & ... & 0.046 & 0.24 \\
MS1910 & 0.246  & ...   & ...         &   ... &  0.33 & 9.6  \\
MS2053 & 0.523 & ...    &   ...      &   ... &   0.119 & 2.60 \\
MS2137 & 0.313 & $4.4^{+0.4}_{-0.4}$ & 960  & ... & 0.0874 & 0.71 \\
MS2318 & 0.130 & 5.1 & ...      & ... & 0.12 & 1.3 \\
AC114  & 0.310 & $8.1^{+1.0}_{-0.9}$&$1649^{+220}_{-220}$ &
                                         0.639   & 0.35   & 13.0 \\
GHO2154 & 0.320 &  ...   &     ...     & 0.721 & 0.0342 & 0.20 \\
PKS0745 & 0.103 & $8.5^{+1.6}_{-1.2}$ & ... & 0.433  & 0.0459  & 0.30 \\
RXJ 1347 & 0.451 & $11.4^{+1.1}_{-1.0}$ & 1235 & ... & 0.24 & 4.2 \\
 & & & & & & \\                                                                  
 \hline                                                             
 & & & & & & \\                                                                  
\end{tabular}
\end{center}
\parbox {7in}
{Data are collected from Wu \& Hammer (1993), 
Le Fevre et al. (1994), 
Kneib \& Soucail (1995), Wu \& Fang (1997),
Allen (1998), Campusano, Kneib \& Hardy  (1998), 
Tyson, Kochanski \& Dell'Antonio (1998) 
and Clowe et al. (1998)}
 \end{table*}

\begin{table*}
\vskip 0.2truein
\begin{center}
\caption{Weak Lensing Cluster Sample}
\vskip 0.2truein
\begin{tabular}{lllllll}

\hline
cluster & $z_{cluster}$ & $T$ (keV) & $\sigma_{gal}$ (km s$^{-1}$) &
$r$(Mpc) & $mass (10^{14}M_{\odot})$ & references \\
\hline
  &  &  &  &  &  &  \\
A1689 & 0.1810 & $9.0^{+0.4}_{-0.3}$ & $1989^{+245}_{-245}$ & 
                      0.20--1.6 & 3.2--14 & Tyson et al. (1995) \\
A2163 & 0.2030 & $13.9^{+0.7}_{-0.5}$ & 1680 &
                      0.09--0.89 & 0.34--4.05 & Squires et al. (1997b) \\
A2218 & 0.1710& $7.0^{+1.0}_{-1.0}$ & $1405^{+163}_{-145}$ & 
                      0.8    & $7.8^{+1.4}_{-1.4}$ & Kneib et al. (1995)\\
      &       &                     &                      &                     
                    0.405  & $2.10^{+0.38}_{-0.38}$ & Smail et al. (1997) \\
      &       &                     &                      &
                      0.25--1.16 & 1.37--11.7 & Squires et al. (1996a) \\
A2263 & 0.208  &    ...    &  ...    &
                      0.87   & $5.0^{+2.5}_{-2.5}$ & Allen (1998) \\
A2390 & 0.2279 & $8.9^{+1.0}_{-0.8}$ & $1093^{+61}_{-61}$ & 
                      0.19-1.20 & 1.59-19.1 & Squires et al. (1996)\\
      &       &                     &                      &
                       0.94 & $10^{+4}_{-4}$ & Allen (1998)  \\
A2744 & 0.308  &   $12.1^{+1.4}_{-1.0}$ & $1950^{+334}_{-334}$ &  
                    0.40 & $3.70^{+0.64}_{-0.64}$ & Smail et al. (1997) \\ 
0957+561 & 0.36 &   ...           & $715^{+130}_{-130}$   &
                       0.06--0.80   & 0.12--1.5 & Fischer et al. (1997a)\\
3C324 & 1.206  &     ...       &     ...                 &
                               0.5 & 6.0 &  Smail et al. (1995)\\
3C295 & 0.46  & 12.6  & $1670^{+364}_{-364}$     &
                  0.40 & $4.70^{+0.76}_{-0.76}$ & Smail et al. (1997) \\ 
3C336 &       &     ...         &     ...                 &
                       0.50 & $4.8^{+1.0}_{-1.0}$ & Bower et al. (1997) \\ 
AC114  & 0.3100 & ...    & $1649^{+220}_{-220}$ & 
                       0.50 & $4.0^{+0.4}_{-0.4}$ & Allen (1998)   \\ 
CL0016 & 0.5545 & $8.0_{-1.0}^{+1.0}$ & $1234^{+128}_{-128}$ & 
                  0.40 & $3.74^{+1.28}_{-1.28}$ & Smail et al. (1997) \\     
CL0024 & 0.3910 &  ...   & $1339_{-233}^{+233}$ &
                       3.0 & 40 & Bonnet et al. (1994) \\
       &        &     &      &
                       0.40 & $2.76^{+0.74}_{-0.74}$ & Smail et al. (1997)\\ 
CL0054 &  0.56      &  ...   &  ...    &
                       0.40 & $3.42^{+1.28}_{-1.28}$ & Smail et al. (1997)\\
CL0303 & 0.0349 &  ...   & $1079_{-235}^{+235}$ & 
                       0.40 & $0.44^{+0.90}_{-0.90}$ & Smail et al. (1997)\\    
CL0412 &  0.51      &   ...  & ...     &
                       0.40 & $0.50^{+0.82}_{-0.50}$ & Smail et al. (1997)\\
CL0939 & 0.4510 & $6.7_{-1.7}^{+1.7}$ & $1081_{-194}^{+194}$  & 
                       0.75 & $6^{+1}_{-1}$ & Seitz et al. (1996)\\ 
       &        &     &      &
                       0.40 & $1.46^{+0.82}_{-0.82}$ & Smail et al. (1997)\\ 
CL1601 & 0.54 &  ...  & 1166    &
                       0.40 & $1.54^{+1.32}_{-1.32}$ & Smail et al. (1997)\\ 
MS1054 & 0.826 & $14.7^{+4.6}_{-3.5}$ & $1643^{+806}_{-343}$  &
                       0.23--2.0  & 0.62--27.3  & Luppino et al. (1997) \\
MS1137 & 0.783 &   ...          &   ...                        &
                       0.18--1.21 & 1.60--4.74  & Clowe et al. (1998)  \\
MS1224 & 0.3255 & $4.3_{-1.0}^{+1.15}$ & $802_{-90}^{+90}$ &  
                       0.96 & 7.0 & Fahlman et al. (1994) \\
MS1358 & 0.3290 & $6.5^{+0.7}_{-0.6}$ & $937^{+54}_{-54}$ & 
                       0.12--1.29 &  0.38--3.65 & Hoekstra et al. (1998)\\
RXJ1347 & 0.4510 & $11.4^{+1.1}_{-1.0}$ & 1235 & 
                       0.24--2.6 & 2.6--30 & Fischer et al. (1997b) \\
RXJ1716 & 0.813 & $6.7^{+2.0}_{-2.0}$ & 1892 &                        
                       0.18--1.11 & 0.95--5.90 & Clowe et al. (1998) \\  
 & & & & & & \\                                                                  
 \hline                                                                        
 & & & & & & \\                                                                  
\end{tabular}
\end{center}
\end{table*}

By searching the literature, we have collected a strong lensing
cluster sample of 38 clusters (Table 1), which contains 48 arc/arclet
images of distant galaxies. We adopt only the published data sets of
the projected cluster masses within the positions of arcs/arclets
and make no attempt as far as possible to extrapolate the original
work. Therefore, a number of fainter and smaller arclets have not been
included in Table 1. Our weak lensing cluster sample consists of 24
clusters, in which the projected radial mass distributions
have been given  for 10 clusters while only one mass measurement
interior to a certain radius is listed for the remaining clusters. 
It is helpful to recall how the cluster mass is computed in the
framework of gravitational lensing. 
In the case of strong lensing, the projected cluster mass
within the position of arc/arclet $r_{arc}$ is often estimated through
\begin{equation}
m_{lens,arc}(r<r_{arc})=\pi r^2_{arc}\Sigma_{crit},
\end{equation}
where $\Sigma_{crit}=(c^2/4\pi G)(D_s/D_dD_{ds})$ is the critical
surface mass density, with $D_d$, $D_s$ and $D_{ds}$ being
the angular diameter distances to the cluster, to the background
galaxy, and from the cluster to the galaxy, respectively. 
Eq.(1) is actually the lensing equation for a cluster lens of 
spherical mass distribution with a negligible small alignment 
parameter for the distant galaxy as compared with $r_{arc}$.  
As for weak lensing, a similar expression to eq.(1) is applied:
\begin{equation}
m_{lens,weak}(r<r_0)=\pi r_0^2 \zeta(r_0) \Sigma_{crit}.
\end{equation}
Here $\zeta$ is defined as (Fahlman et al. 1994)
\begin{eqnarray}
\zeta(r_0) & =   2\left(1-r_0^2/r_{max}^2\right)^{-1}
\int_{r_0}^{r_{max}}\langle\gamma_T\rangle d\ln r \nonumber \\
  &  =\overline{\sigma}(r<r_0)-\overline{\sigma}(r_0<r<r_{max}),
\end{eqnarray}
in which $\langle\gamma_T\rangle$ is the mean tangential shear around
a circular path of radius $r$ introduced by the gravitational potential of
the cluster, 
$\overline{\sigma}(<r_0)$ and $\overline{\sigma}(r_0<r<r_{max})$
represent the mean surface mass density interior to $r_0$ and
in the annulus $r_0<r<r_{max}$, respectively. Therefore, eq.(2)
provides a low bound on the projected cluster mass within $r_0$.

In order to reconstruct the cluster mass  $m_{lens,arc}(r<r_{arc})$ or
$m_{lens,weak}(r<r_0)$, one needs also to know the redshift of the arc/arclet
or the spatial distribution of background galaxies.  
Spectroscopic measurements have been made for about $\sim2/5$ of the
arcs/arclets in Table 1. When the data of redshifts are not available,
a redshift of $z_s=0.8$ is often used in the estimate of
$\Sigma_{crit}$, except for the arcs in clusters MS1137+66
at $z=0.783$ where $z_s=1.5$ is adopted in Table 1.   
Such an assumption is  justified by the recent work of 
Ebbels et al. (1998), who obtained the spectroscopic identifications 
of 18 arclike images behind A2218 and found a mean redshift of 
$\langle z_s\rangle=0.8$--1 at $R\sim25.5$. Nevertheless,
placing the background galaxies at $z_s=2$ instead of $z_s=0.8$ 
would reduce the value of $\Sigma_{crit}$ by a factor of only $1.4$
for a mean cluster redshift of $\langle z_d\rangle\approx0.3$.    
Weak lensing measurements involve a great number of faint and distant 
galaxies behind foreground clusters, and the spectroscopic measurements 
of all the population of galaxies in the fields seem to be impossible 
at present. Different calibrations have thus been applied by different authors 
in the determinations of $\Sigma_{crit}$.
For clusters at relatively lower redshifts of $z_d\approx0.3$, 
it appears to be plausible to assume the mean redshift of the background
galaxies to be at $z_s\sim$0.8--$1.2$ when combined with the existing
work of deep galaxy surveys. However,  it should be noted that 
the big uncertainty of up to a factor of $\sim5$ arising from
the unknown redshift distributions of background galaxies may occur in
the evaluation of $\Sigma_{crit}$ for
high-$z$ clusters at $z_d\approx0.8$ (e.g. Luppino \& Kaiser 1997).

Before we turn to the comparison of strong lensing, weak lensing and
X-ray/optical determined cluster masses, we display in Fig.1 the
relationship between the velocity dispersion ($\sigma_{gal}$) of
galaxies and the temperature ($T$) of X-ray emitting gas for
20 lensing clusters in Table 1 and Table 2, for which
both $\sigma_{gal}$ and $T$ are available in literature.
It is believed that the $\sigma_{gal}-T$ relationship can provide
a straightforward yet robust test for the dynamical properties of
clusters of galaxies (Cavaliere \& Fusco-Femiano 1976).  
Employing the least-square fit of a power-law to the data of Fig.1 yields 
$(\sigma_{gal}/$km s$^{-1})=10^{2.57\pm0.13}(kT/$keV)$^{0.59\pm0.14}$.
Regardless of its large error bars, this relationship is essentially 
identical with the one for the nearby clusters (Girardi et al. 1996 
and references therein), and is also consistent with the isothermal 
hydrostatic scenario of $\sigma_{gal}\sim T^{0.5}$.
Based on the $\sigma_{gal}-T$ relationship alone,
we may conclude that no significant dynamical and cosmic evolution
has been detected  for those massive lensing clusters
within intermediate redshift $\langle z_d\rangle\approx 0.3$.
It is interesting to recall that similar conclusions have been 
reached by numerous recent studies on X-ray clusters.
For instances, the cluster number counts 
exhibit no evolutionary tendency at least to
redshift of as high as $z\sim0.8$ (Rosati et al. 1998), and
no significant differences in the X-ray luminosity-temperature 
relationship and the velocity dispersion-temperature relationship 
between low-redshift and high-redshift clusters are seen
(e.g. Mushotzky \& Scharf 1997). 
In particular, the distribution of core radius of the intracluster
gas of nearby clusters accords with that of distant clusters
($z>0.4$) (Vikhlinin et al. 1998). So,
these arguments may eventually support the hypothesis that overall,
both galaxies and gas are the tracers of the depth and shape of 
the underlying gravitational potential of cluster, in despite of  
the presence of substructures and merging activities on small scales.

\section{Comparisons of different mass estimates}

\subsection{Strong and weak lensing}

Ten weak lensing clusters in Table 2 also contain arclike images  of 
background galaxies in their central cores. Comparison of strong and 
weak lensing determined cluster masses among these clusters are 
straightforward  and shown in Fig.2.  
Except for A1689 where the absolute mass calibration was
made using the giant arcs which trace the Einstein radius 
(Tyson \& Fischer 1995),
the rest strong lensing events seem to yield  larger 
cluster masses than the weak lensing measurements  when the two methods
become to be comparable with each other at the central regimes. 
In Fig.3 we display the strong/weak lensing 
determined cluster masses at different radii utilizing  
all the data sets in Table 1 and Table 2. 
Although such a comparison sounds less serious in the sense that
they are different clusters, it is by no means of physical insignificance.
Indeed, the majority of clusters of galaxies that are capable of 
magnifying and distorting the images of background galaxies 
are very rich clusters at intermediate redshift. 
They are representative of the (most) massive clusters at 
$\langle z\rangle\sim0.3$. Statistically, their gross dynamical properties 
should look very similar.  It is remarkable that 
the data points in Fig.3 are clearly separated into two parties: 
the strong lensing determined cluster masses systematically and 
significantly exceed the weak lensing values. Actually, the strong lensing
data appear in the plot as if they were the upper limits to the
weak lensing measurements. Alternatively, the cluster masses given by 
the arcs/arclets with and without confirmed redshifts show no
significant differences. The very small dispersion in the strong lensing
results arises from the fact that the critical surface mass density
parameter $\Sigma_{crit}$ remains roughly unchanged
for the known lensing systems which have a mean cluster redshift 
of $\langle z_d\rangle\approx0.3$ and a mean source redshift of
$\langle z_s\rangle\approx0.8$. Consequently, the ``mass density profile''
derived from the  strong lensing events varies as $r^{-2}$.

\subsection{Lensing and X-ray measurements}

X-ray observations have been made for most of the lensing clusters.
Under the assumption that the hot diffuse gas is in hydrostatic 
equilibrium with the underlying gravitational potentials of clusters,
one can easily obtain the X-ray cluster masses provided that the gas and
temperature radial profiles are well determined.  Adopting the conventional
isothermal $\beta$ model for gas distribution which is characterized 
by the core radius $r_{xray,c}$, the index $\beta_{fit}$ 
and the temperature $T$, we can write out the
projected X-ray cluster mass within radius $r$ to be (Wu 1994)
\begin{equation}
m_{xray}=1.13\times10^{13}\beta_{fit}\tilde{m}(r)
            \left(\frac{r_{xray,c}}{0.1\;{\rm Mpc}}\right)
            \left(\frac{kT}{1\;{\rm keV}}\right)\; M_{\odot},
\end{equation}
where 
\begin{eqnarray*}
\tilde{m}(r)=& \frac{(R/r_{xray,c})^3}{(R/r_{xray,c})^2+1} 
               \;\;\;\;\;\;\;\;\;\;\;\;\;\;\;\;\; \\
           & -\int_{r/r_{xray,c}}^{R/r_{xray,c}}x\sqrt{x^2-(r/r_{xray,c})^2}   
             \frac{3+x^2}{(1+x^2)^2}dx,
\end{eqnarray*}
and $R$ is the physical radius of the cluster and will be taken to be
$R=3$ Mpc in the actual computation. Our conclusion is unaffected 
by this choice.

We have computed the projected X-ray cluster mass $m_{xray}$ interior to
the position of each arclike image $r_{arc}$
or the corresponding radius of 
each weak lensing measurement $r_{0}$
for the 21/13 strong/weak lensing clusters
of known temperatures listed in Table 1 and Table 2, and 
the resulting $m_{xray}$ versus $m_{lens}$ are plotted in Fig.4 
for $\beta_{fit}=2/3$ and a mean core radius of 
$\langle r_{xray,c}\rangle=0.25$ Mpc. 
It turns out that although
both the strong and weak lensing determined cluster masses 
$m_{lens,arc}$ and $m_{lens,weak}$
show a good correlation with the X-ray masses $m_{xray}$,
their amplitudes  are very different:
$m_{lens,arc}/m_{xray}=3.23\pm1.21$ and 
$m_{lens,weak}/m_{xray}=0.97\pm0.44$. The scatter in the fit of
$m_{lens,weak}/m_{xray}$ can be reduced when cluster A2163 is
excluded (Fig.4b) (Note that A2163 is one of the hottest clusters 
known so far).  That is to say, there is a significant discrepancy 
between the strong lensing derived cluster masses and the X-ray 
cluster masses, while an excellent agreement between the weak lensing 
results and the X-ray masses is found. The only way to reconcile 
$m_{lens,arc}$ with $m_{xray}$ is to adopt
a considerably small core radius for the gas profile. Fig.5 demonstrates
another plot of $m_{lens}$ against $m_{xray}$ by reducing $r_{xray,c}$ to
$0.025$ Mpc. In this case, the ratios of 
$m_{lens,arc}$ and $m_{lens,weak}$ to $m_{xray}$ read 
$1.42\pm0.87$ and $0.91\pm0.35$, respectively. 
Yet, this does not relax the disagreement
between $m_{lens,arc}$ and $m_{lens,weak}$. Rather, the employment of
a smaller core radius leads to a significant increase of the X-ray cluster
mass estimate at small radius. Recall that the crucial point behind
the remarkable agreement of the strong lensing and X-ray determined masses
for the cooling flow clusters reported by Allen (1998) is the
small core radii of $r_{xray,c}\approx40$--75 kpc, in contrast to 
$r_{xray,c}\approx250$ kpc for the non-cooling flow clusters.

\subsection{Velocity dispersion of galaxies as the tracer of 
            cluster potential}

If galaxies trace the underlying gravitational potential of cluster,  
their velocity dispersion $\sigma_{gal}$ would be
a good indicator of dark matter. A strict way to derive the
dynamical mass of cluster from the distributions of galaxy population
and their velocity dispersion is to work with the Jeans equation.
Here, we utilize an alternative approach to the issue. 
We adopt the so-called softened isothermal sphere model 
with a core radius $r_{dark,c}$
for the total mass distribution of cluster, which is 
characterized by the velocity dispersion of galaxies.
The original motivation was to
examine whether $\sigma_{gal}$ provides a proper cluster mass estimate
(Wu \& Fang 1997). The projected cluster mass within radius $r$ 
is simply
\begin{equation}
m_{opt}(<r)=\frac{\pi \sigma_{gal}^2}{G}
        \left(\sqrt{r^2+r_{dark,c}^2}-r_{dark,c}\right).
\end{equation}
There are 21/18 clusters in Table 1/2 whose velocity dispersions
are observationally determined. We compute their total masses
in terms of eq.(5) at the corresponding positions  of arclike
images or weak lensing measurements. We first adopt a core
radius of $r_{dark,c}=0.25$ Mpc, in accord with  the 
distributions of luminous matter (galaxies and gas) in cluster,
and the resulting $m_{opt}$ are shown in Fig.6.
It turns out that the weak lensing determined cluster masses
$m_{lens,weak}$ are in fairly good agreement with $m_{opt}$ despite of  
the large scatters: $m_{lens,weak}/m_{opt}=1.08\pm0.70$,
whereas the strong lensing results $m_{lens,arc}$ 
depart apparently from the expectation of $m_{lens,arc}=m_{opt}$ with 
$m_{lens,arc}/m_{opt}=6.07\pm3.98$. 
Motivated by the argument that 
the dark matter profile is sharply peaked at the cluster center 
relative to the luminous matter distribution 
(e.g. Hammer 1991; Wu \& Hammer 1993; Durret et al. 1994; etc), 
we also present in Fig.6 the results $m_{opt}$ for a smaller
core radius of $r_{dark,c}=0.025$ Mpc.
This indeed reduces the difference between $m_{lens,arc}$ and $m_{opt}$,
but simultaneously breaks down the accordance of $m_{lens,weak}$ and 
$m_{opt}$. The mean ratios of strong and weak lensing results 
to the isothermal sphere model determined cluster masses are 
now $m_{lens,arc}/m_{opt}=1.44\pm0.97$ and  
$m_{lens,weak}/m_{opt}=0.63\pm0.35$, respectively. 
The scatter in $m_{lens,weak}/m_{opt}$ can be significantly reduced if 
the data sets of A2163 and RXJ1716 are excluded.

A more reasonable way to estimate the dynamical mass of cluster in terms
of velocity dispersion of galaxies is to employ the virial theorem:
$M_{vir}=3\sigma_{gal}^2r_v/G$, which measures the total cluster mass enclosed
within a sphere of the so-called virial radius $r_v$. A direct comparison
of the virial and lensing cluster masses is somewhat difficult because 
the virial radius $r_v$ is usually much larger than the size which can be
reached by the current gravitational lensing techniques. 
We display in Fig.7 the cluster masses given by strong/weak lensing and 
virial theorem for 10 clusters with the available data of $M_{vir}$
in literature. The projected cluster mass derived from 
gravitational lensing would approach to the 3-D virial mass at
radius as large as $r_v$. However, the fact that the cluster regimes 
probed by the two methods show no overlaps can only lead us to 
arrive at the conclusion that these two independent methods seem to
provide a consistent radial matter distribution of cluster.

\section{Discussion}

\subsection{Strong lensing: overestimates cluster mass ?}

The consistence of the $\sigma_{gal}$-$T$ relationship of the 
lensing clusters with  $\sigma_{gal}\propto T^{0.5}$ 
expected under the hypothesis of 
isothermal and hydrostatic equilibrium    
suggests that the gas and galaxies are good tracers of 
underlying gravitational potential of the clusters.  
Indeed, a number of recent studies have shown that clusters
of galaxies suffer from little evolution, and their dynamical
properties have remained almost unchanged since $z\sim0.8$ 
(Mushotzky, \& Scharf 1977; Bahcall, Fan, \& Cen 1997; 
Henry 1997; Rosati et al. 1998;  Vikhlinin et al. 1988; etc.).
This essentially justifies the employment of the Jeans equation
for X-ray emitting gas in the lensing clusters at intermediate
redshifts.   On the other hand, the weak lensing probes
the gravitational potential fields of the clusters in a completely
different way. So, the excellent agreement 
between weak lensing and X-ray determined cluster masses
serves as another convincing evidence for the lensing clusters 
being the dynamically relaxed systems.  

A further examination of the mass determinations with eq.(1) 
tells us that there are no free parameters with which one can play 
when the redshifts of the lensing clusters and of the arclike images
are observationally measured. It is unlikely that 
the replacement of the simple spherical lens model by a more
sophisticated one can make a significant difference 
[see Allen (1998) for a detailed discussion; and references therein). 
Therefore, we are forced to accept the fact that the present modeling
of strong lensing event does provide a reliable cluster mass estimate. 

However, from the comparisons of different cluster mass estimators
in the above section, it is apparent  
that the cluster mass revealed by strong lensing
exceeds that inferred from weak lensing and X-ray method by a factor
of $\sim2$ - $4$.  If we choose to trust the strong lensing result,
we may need to break down the fairly good agreement between 
the weak lensing and X-ray derived cluster masses, and 
work out a mechanism which can account for the mass discrepancy 
between the strong lensing and the latter two methods.
There are quite a number of mechanisms such as projection effect 
and asymmetrical mass distributions which may  
result in  an overestimate of cluster mass from arclike images. 
But, these mechanisms will simultaneously influence the weak lensing 
results.  In the following discussion we will focus on whether 
weak lensing and X-ray analysis give rise to an underestimate of 
the gravitating masses of clusters.

\subsection{Weak lensing: underestimates cluster mass ?}

In principle, the weak lensing method eq.(2) and (3) always provides 
a lower bound on the cluster mass $m_{lens,weak}(<r_0)$ interior to 
radius $r_0$. So, the total mass within $r_0$ could be considerably
underestimated unless the outer control annulus $r_{max}$ is 
set to be sufficiently large and/or the true mass density profile drops 
sharply along outward radius. In the current measurements 
of weak lensing effects, 
a value of $r_{max}\sim1$ -- 2 Mpc is often adopted due to the
limited data coverages.  If cluster exhibits rather a 
flat surface mass density, e.g. a power-law of $\propto r^{-\gamma}$
with index $\gamma<1$, it is not impossible that 
the true cluster masses can be underestimated by a factor of as large as 
$\sim2$, according to the formalism of Kaiser et al. (1994):
$\overline{\sigma}/\zeta=(a^2-1)/(a^2-a^{2-\gamma})$ where $a=r_0/r_{max}$.

Another crucial point is relevant to the application of weak lensing
inversion technique to the central region of cluster close to 
the Einstein radius, where the actual comparison of
strong lensing and weak lensing determined cluster
masses becomes possible. However, because the giant arc traces approximately
the Einstein radius and because the weak lensing method can be 
marginally applicable to such a small radius, large uncertainties 
could be introduced into the shear measurements  within
the positions of arcs/arclets (e.g. Seitz \& Schneider 1995). 
This even does not account for the possible
contamination of central cluster galaxies due to the small number of 
background faint galaxies near the Einstein radius. From Fig.2 
it is easily recognized that the innermost radii of the weak 
lensing measurements are always larger than or equal to the arc radii.
Therefore, one cannot exclude the possibility that 
the disagreement of the strong and weak inferred cluster
masses  arises simply from the inappropriate application of
the week lensing inversion technique to the central regimes of clusters,
although it has been shown that the cluster mass 
is only slightly underestimated at the cluster center if 
a general method for the nonlinear
cluster mass reconstruction is used 
(Kaiser 1995; Schneider \& Seitz 1995; Seitz \& Schneider 1995).

\subsection{X-ray analysis: underestimates cluster mass ?}

An accurate estimate of cluster mass from the X-ray measurement 
depends on our understanding of the gas distribution and 
its dynamical state in cluster, which is closely connected to
how significant the X-ray emitting gas deviates from   
the state of hydrostatic equipartition 
with the gravitational potential of the whole cluster. 
This argument is twofold: (1)The complex dynamical 
activities in cluster may only take place locally and on small scales 
such as the substructure merging and irregular temperature patterns, 
while the cluster as a whole can be regarded as a violently relaxed 
system; (2)Cluster is still in forming stage and thereby, cannot be
modeled as a virialized system at all.  

In the first circumstance, the Jeans equation may be safely 
applied to large radius but fails at small scales. Such a dynamical
model can account for the excellent agreement between weak lensing 
and X-ray cluster masses at relatively large radius, and the mass discrepancy 
between X-ray and strong lensing measurements within the central cores 
of clusters.  Further evidences for this scenario come from 
the overall $\sigma_{gal}$-$T$ relationship for
the lensing clusters (Fig.1) and the recent observations of cluster
abundances and other dynamical properties at redshift out to $z\sim0.8$
(Mushotzky, \& Scharf 1997; Bahcall et al. 1997; Rosati et al. 1998; etc.),
i.e., the significant cosmic and dynamical evolutions do not play a 
dominant role for clusters of galaxies as a whole since $z\sim0.8$.
There are a number of mechanisms that could affect                     
the determinations of cluster masses on small scales 
based on the X-ray observations, among which
the presence or lack of the cooling flows inside the cores of clusters 
is likely to be the major source of uncertainties (Allen 1998).  
If the core radius of the X-ray emitting gas profile 
is overestimated due to the contamination of these central 
dynamical activities, we can indeed reconcile the strong 
lensing determined cluster masses with the X-ray measurements 
by substantially reducing the X-ray core radius (Fig.5).   
Recent analyses by Markevitch (1997) and Girardi et al. (1997b) for 
A1689 and A2218, 
based on the high resolution X-ray/optical images, have revealed that 
the local dynamical activities like the ongoing subcluster 
mergers in the core regions probably make the hydrostatic mass estimate 
inapplicable. Additional supports come from the 
numerical simulations of cluster formation and evolution. For instances,
Bartelmann \& Steinmetz (1996) showed that utilizing a $\beta$ model 
for a dynamical active cluster may lead to an average  
underestimate of the true cluster mass by a factor of $\sim40\%$, and
strong lensing preferentially selects clusters that are
dynamically more active than the average.
From an extensive analysis of the $\beta$ model
as a cluster mass estimator, Evrard, Metzler \& Navarro (1996) 
concluded that while the mass estimates based on 
the hydrostatic, isothermal $\beta$ model are remarkably accurate,
the ratio of the estimated cluster mass in terms of the $\beta$ model to
the true cluster mass increases with increasing radius, with the true
cluster mass being underestimated by the $\beta$ model toward 
cluster center when $\Omega_0=1$, and merging rates are increased 
at low redshift. This is fairly consistent with our speculation (1).

The second possibility is of interest when a large $\Omega_m$ universe
(e.g. $\Omega_m=1$) is preferred, in which clusters of galaxies formed
very recently and may still be in a stage of dramatic dynamical evolution.
This may essentially invalidate the application of the Jeans
equation without considering the infall motions of the materials inside and
around the clusters. The coincidence 
between the weak lensing and X-ray determined cluster masses (see Fig.4)
does not ensure that both methods would not result in 
an underestimate of total cluster masses.  It is somewhat unfortunate 
if we have to accept this scenario, though there have been no 
convincing observations so far to confirm this argument.

\subsection{Velocity dispersion of galaxies: a good indicator of cluster ?}

Unlike the diffuse X-ray gas, cluster galaxies are unaffected by the presence 
of the (non)cooling flows and the nonthermal pressure such as
magnetic field and therefore, are probably a better indicator of 
mass distributions of clusters. However,
a precise measurement of the galaxy velocity dispersion
to large radius from the cluster center is difficult. Employment of
the Jeans equation for galaxies among the ensemble clusters including 
lensing ones at intermediate redshifts have become possible only recently
(Carlberg, Yee, \& Ellingson 1997). On the other hand, 
the virial and lensing cluster 
masses are basically consistent with each other (see Fig.7), though
the two methods probe very different regions of clusters.

Attempts have been made to fit the lensing derived 
mass profile with either a singular or softened isothermal sphere model
for many individual lensing clusters (e.g. Fahlman et al. 1994; 
Tyson \& Fischer 1995; Squires et al. 1997a,b; Luppino \& Kaiser 1997;  
Fischer \& Tyson 1998; Tyson, Kochanski, \& Dell'Antonio 1998; etc),
which have nevertheless yielded a controversial result. From our 
statistical study (see Fig.6), an uncomfortably small core radius of $\sim0$
may be required for an isothermal dark matter distribution 
if the strong lensing data alone are used, while
the weak lensing results allow the core size of dark matter profile 
to be as large as the one for luminous matter distributions. 
A combination of the strong and weak lensing analyses 
suggests that an isothermal sphere model with  a compact but non-zero core  
and characterized by the observed velocity dispersion
of galaxies $\sigma_{gal}$ is likely to provide a good description of 
total mass distribution of cluster. This is consistent with
the first high resolution mass map of CL0024+1654 through parameter 
inversion of the multiple images of a background galaxy (Tyson et al.
1998), which has detected the presence of a soft core of $\sim0.066$ Mpc. 
Alternatively, if the true mass distribution of cluster is close to an 
isothermal sphere model, the statistical agreement of the lensing
measured cluster masses with those expected from the isothermal
model indicates that there is no strong bias between the velocity of the
dark matter particles and of the galaxies. In other words, velocity
dispersion of galaxies seems to be a good indicator of the gravitating 
mass of cluster.

\section{Conclusions}

Statistical comparisons of different mass estimators among 
lensing clusters based on the published data in literature
have revealed the following features: (1)Strong gravitational lensing,
which is believed to the most reliable mass estimator, 
gives rise to a systematically larger projected cluster within the 
position of arclike image than those derived from the weak lensing 
technique and the X-ray measurements; (2)There is an excellent
agreement between the weak lensing and X-ray derived cluster masses;
(3)An isothermal sphere with a compact core radius, which is
characterized by the velocity dispersion of galaxies, 
provides a good description of total mass distribution of cluster. 

It is possible to reconcile the different cluster mass estimators
under the following way: both the weak lensing technique and the 
X-ray measurements based on the hydrostatic equilibrium hypothesis 
are inappropriate to  probe of the matter distributions in the
central regions of clusters, which may underestimate
the gravitating masses enclosed within the cluster cores
by a factor of $2$--$4$ as compared with the strong lensing method.  
Nevertheless, these cluster mass estimators may be safely applied on
large scales outside the core radius, which is supported by 
the studies of  dynamical properties of clusters
and also the excellent agreement between the weak lensing and X-ray measured
cluster masses. Basically, a smaller core radius of $r_{dark,c}\sim$
a few ten kpc is needed for both dark and luminous matter profiles 
in order to explain the detected  mass discrepancy. 
Very likely, it is the local
dynamical activities that lead the dark matter profile to appear more
peaked than the luminous matter distributions at the centers of clusters. 
Overall, the light profile traces the dark matter distribution 
in cluster, and the velocity dispersion of galaxies and the temperature of
gas are both good indicators of the underlying gravitational potentials
of the whole clusters.

The cosmological implications of this work are as follows: 
The mass-to-light ratio $M/L$ and 
the baryon fraction $f_b$ revealed by the current optical
and X-ray observations of clusters are indeed reliable indicators for
the overall matter composition of clusters and thereby the universe,
though these quantities  may have rather a large uncertainty 
on small scales of $\sim100$ kpc.  
As a result, we have to accept  a low mass density universe of
$\Omega_m\approx0.1$-$0.4$, as suggested by 
the two independent measurements of the cluster matter
compositions, the ratio $M/L$  [(dark+luminous)/luminous]
(e.g. Bahcall, Lubin, \& Dorman, 1995) 
and the baryon fraction $f_b$ [baryon/(baryon+nonbaryon)]
(e.g. White et al. 1993). If so, 
we should be aware that the fraction of baryonic
matter in the universe is not small at all: $\sim20$-$30\%$ of the
matter in the universe is luckily visible !


\section*{Acknowledgments}

We gratefully acknowledge the referee, Vincent Eke, for 
valuable comments and suggestions. 
This work was supported by the National Science Council of Taiwan,
under Grant No. NSC87-2816-M008-010L and NSC87-2112-M008-009, and 
the National Science Foundation of China, under Grant No. 1972531.

\begin{figure*}
\centerline{\hspace{3cm}\psfig{figure=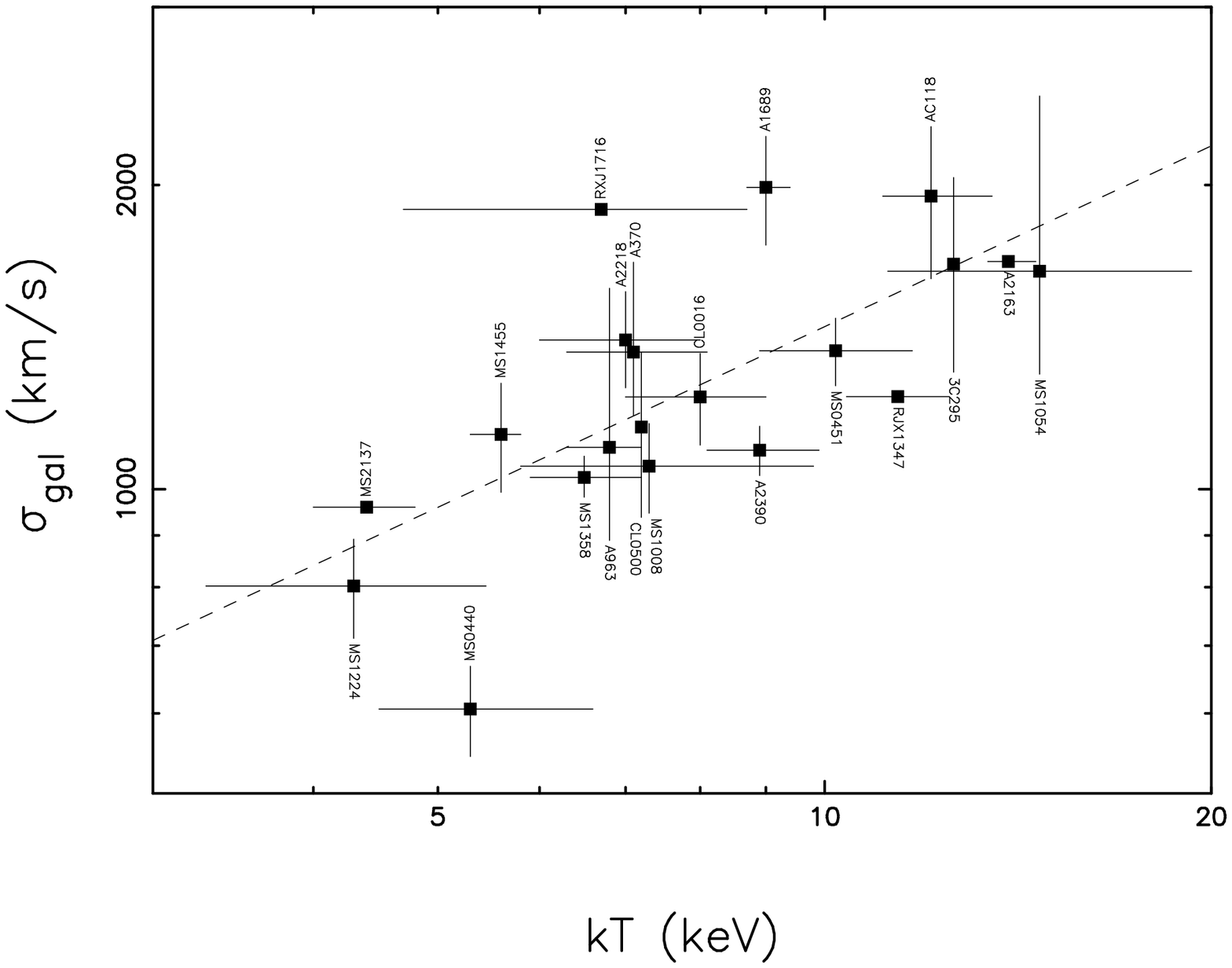,width=0.9\textwidth,angle=270}}
\caption{The $\sigma_{gal}$-$T$ relationship of the 20 lensing clusters
in Table 1 and Table 2, for which both velocity dispersion $\sigma_{gal}$
and temperature $T$ are observationally determined. The dashed line 
is the best-fit to the data: 
$(\sigma_{gal}/{\rm km\;s^{-1}})=
10^{2.57\pm0.13}(kT/{\rm keV})^{0.59\pm0.14}$. 
}
\end{figure*}

\begin{figure*}
\centerline{\hspace{3cm}\psfig{figure=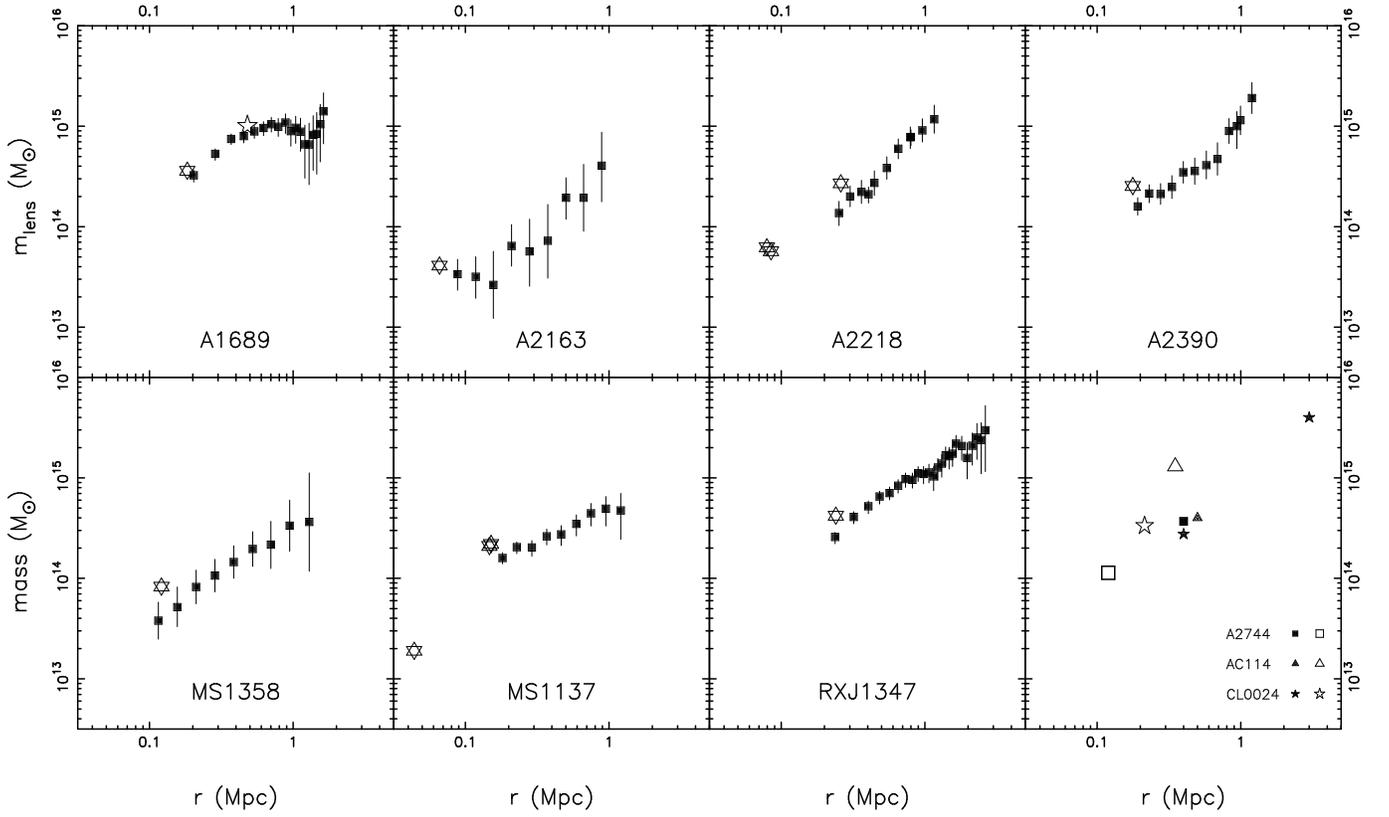,width=0.9\textwidth,angle=270}}
\caption{Comparison of the strong and weak lensing determined cluster
masses among individual clusters. The open and filled symbols 
represent the strong and weak lensing results, respectively. 
Note that the giant arc was used for  
the calibration of the weak lensing measurements in A1689, for which  
we have also displayed the recent result (open asterisk) 
obtained from the measurement of the deficit of red galaxy 
population behind A1689 for comparison (Taylor et al. 1998).
The uncertainties of the strong lensing results are not shown, which turn
to be quite small when 
the positions and redshifts of the arcs are observationally determined.    
}
\end{figure*}

\begin{figure*}
\centerline{\hspace{3cm}\psfig{figure=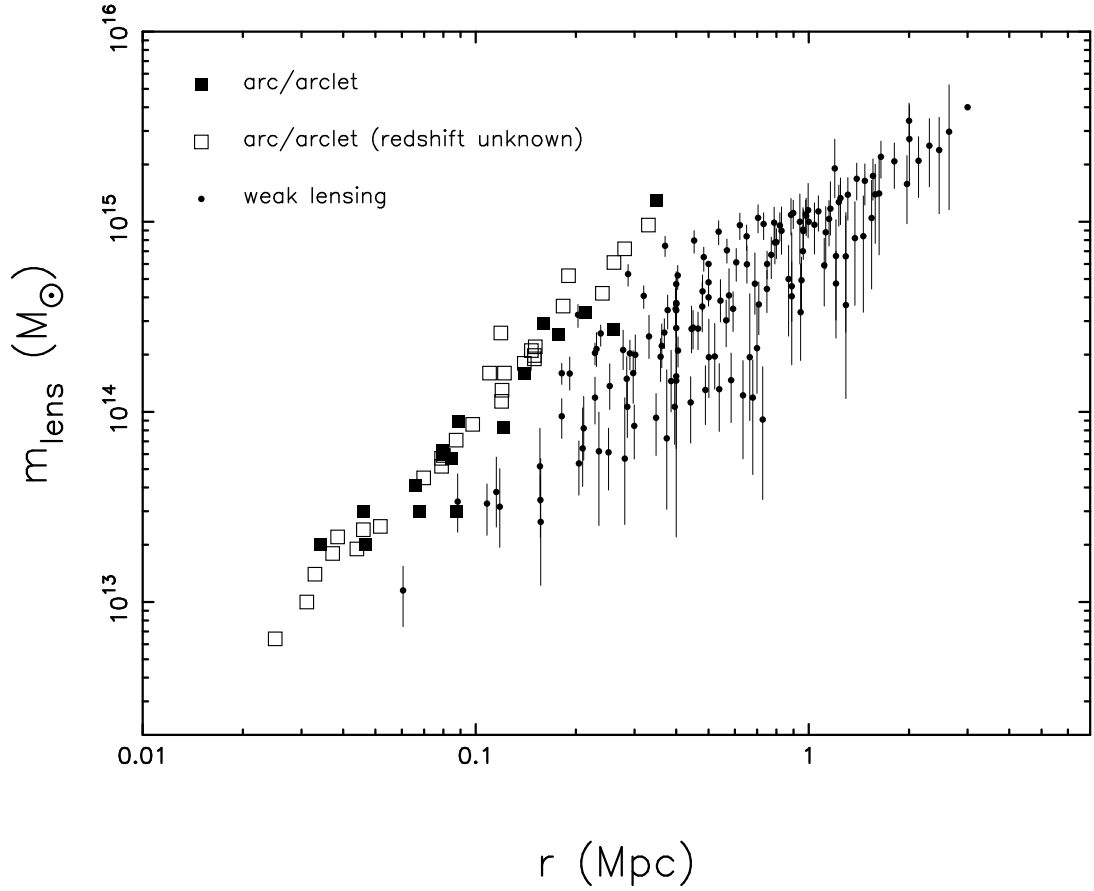,width=0.9\textwidth,angle=270}}
\caption{The strong and weak lensing measured cluster masses are
plotted against the cluster radii for all the data sets in Table 1
and Table 2. It appears that the gravitating masses revealed by the
arclike images systematically exceed those derived from the
inversion of the weakly distorted images of background galaxies 
around clusters. 
}
\end{figure*}

\begin{figure*}
\centerline{\hspace{1.cm}\psfig{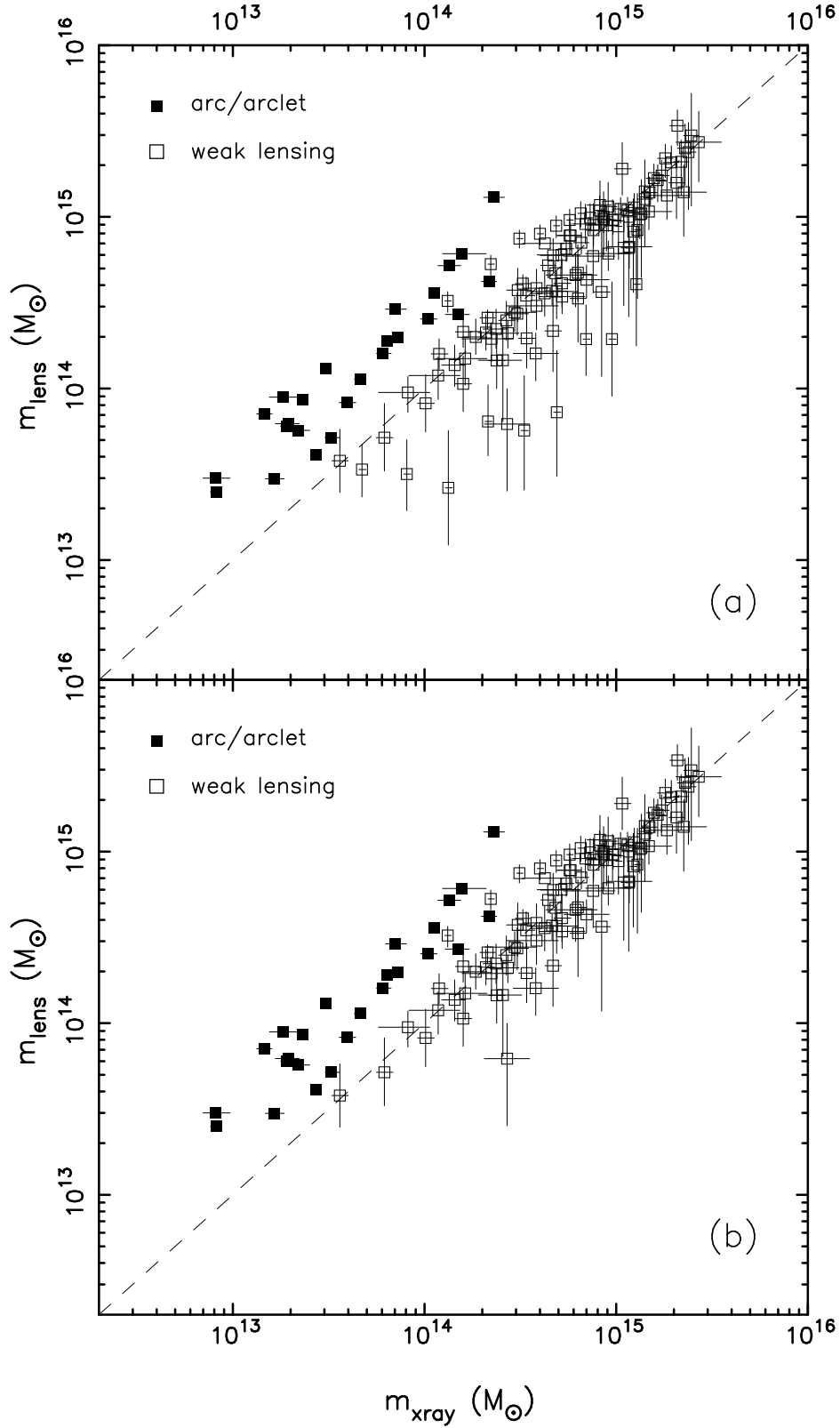}}
\caption{Gravitational lensing determined cluster masses $m_{lens}$ versus the
hydrostatic masses $m_{xray}$ given by the X-ray diffuse gas. 
Only those clusters in Table 1 and Table 2 whose temperatures are 
spectroscopically measured are shown.  A conventional isothermal 
$\beta$ model with $\beta_{fit}=2/3$ and core radius $r_{xray,c}=0.25$ Mpc 
is adopted for the gas distribution. It is remarkable that $m_{xray}$
agree essentially with the weak lensing results (a), while an excellent
agreement is reached when cluster A2163 is excluded (b). 
The horizontal error bars only reflect the uncertainties in the measurements
of temperature. The dashed line
is not a fit to the data but assumed that $m_{lens}=m_{xray}$.   
}
\end{figure*}

\begin{figure*}
\centerline{\hspace{1cm}\psfig{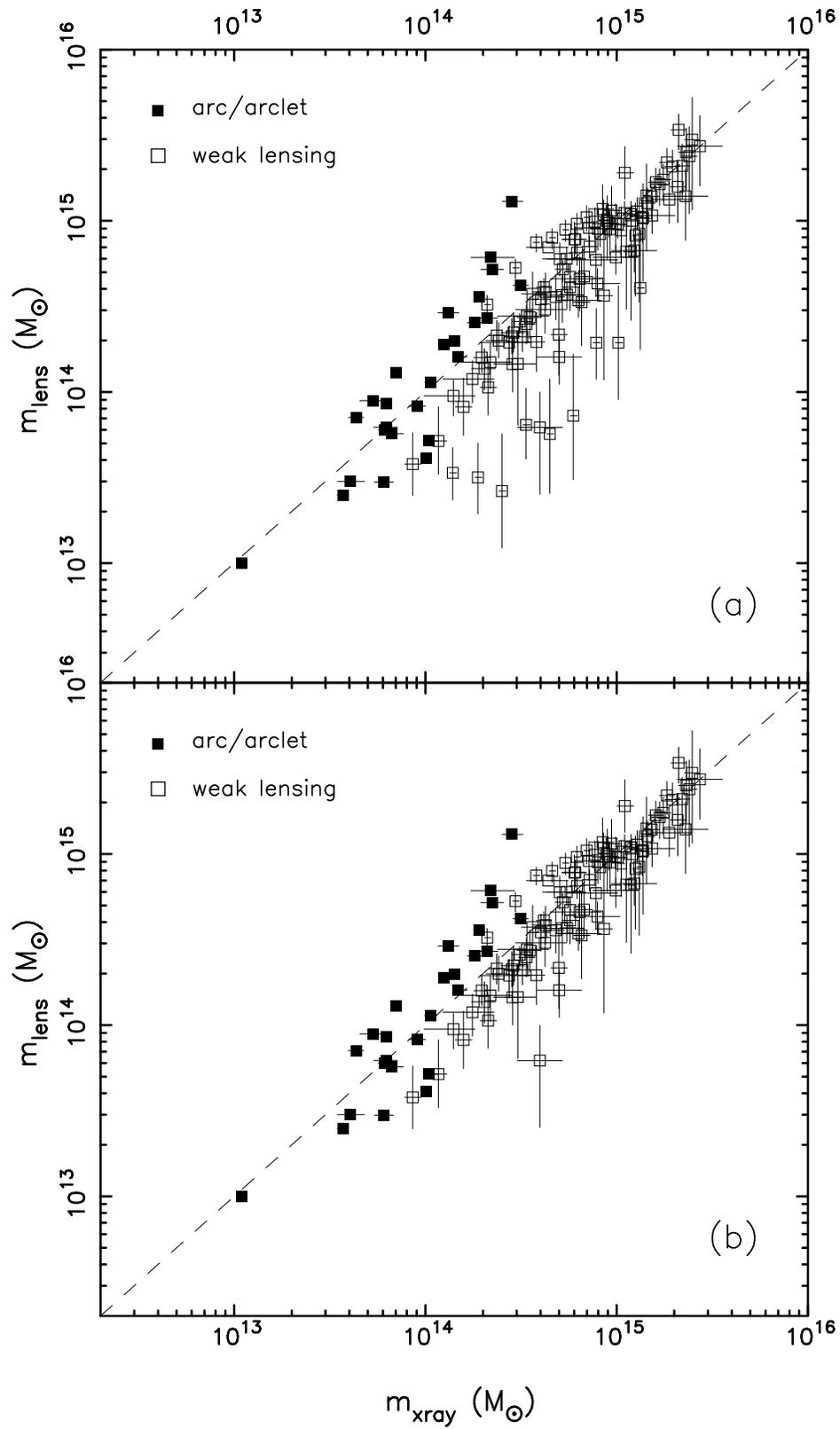}}
\caption{The same as Fig.4 but for a much smaller core radius of
$r_{xray,c}=0.025$ Mpc in the gas profile.  
}
\end{figure*}

\begin{figure*}
\centerline{\hspace{1cm}
\psfig{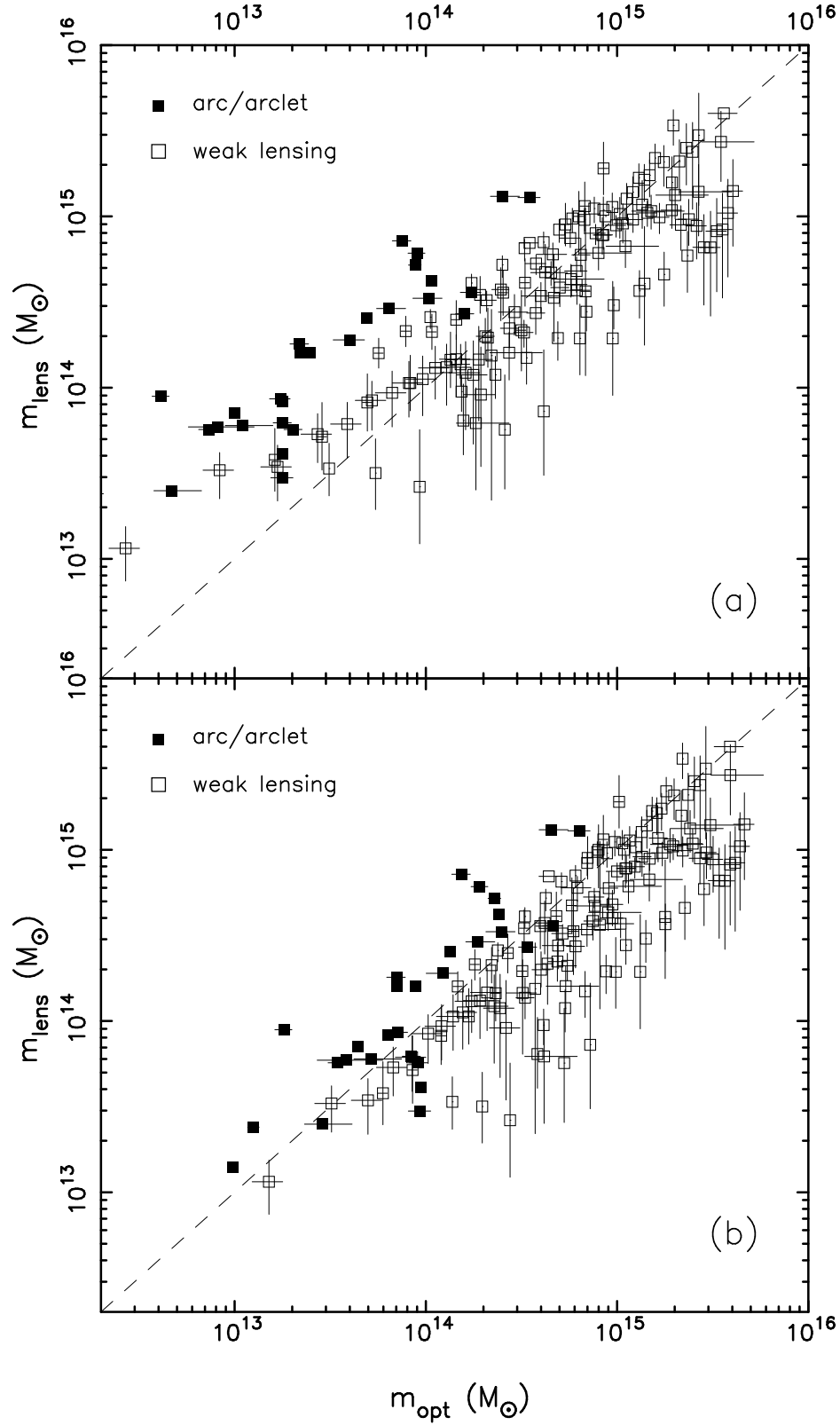}}
\caption{Gravitational lensing determined cluster masses $m_{lens}$
are plotted against the theoretically expected results $m_{opt}$
from a softened  isothermal sphere model for the total mass distributions of
clusters, which is characterized by the velocity dispersion 
$\sigma_{gal}$ of galaxies and a core radius $r_{dark,c}$. 
(a) $r_{dark,c}=0.25$ Mpc and (b) $r_{dark,c}=0.025$ Mpc. 
Only the clusters whose $\sigma_{gal}$  are observationally measured are 
shown, and the horizontal error bars represent the uncertainties in 
$\sigma_{gal}$. The dashed line
is not a fit to the data but assumed that $m_{lens}=m_{opt}$.
The dispersions in $m_{lens}$ can be considerably reduced if the weak
lensing data of A2163 and RXJ1716 are excluded. 
}
\end{figure*}

\begin{figure*}
\centerline{\hspace{3cm}\psfig{figure=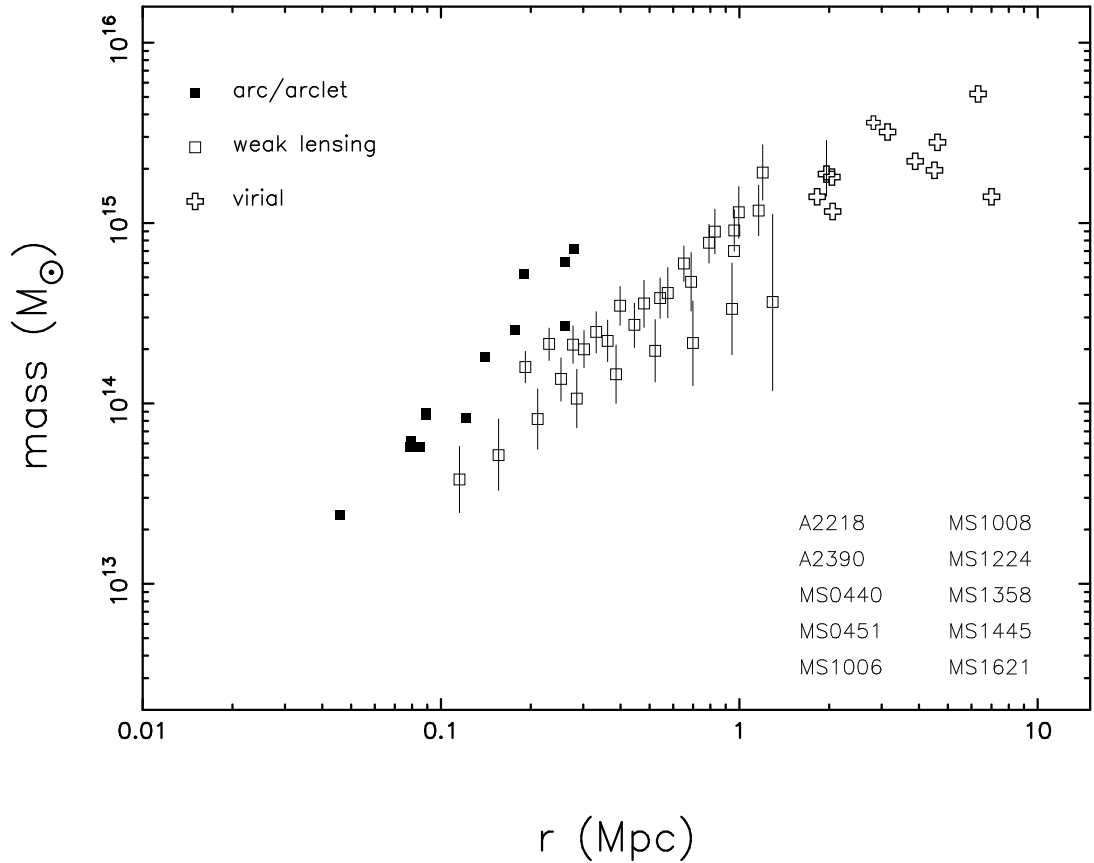,width=0.9\textwidth,angle=270}}
\caption{A comparison of the cluster masses derived from 
gravitational lensing and virial theorem.  
At sufficiently large radius, the deviation of the projected 
masses from the spatial (virial) values should become negligible. 
According to the present data sets, it is unlikely that 
there are significant differences between the two mass estimators.
The virial masses are from Carlberg et al. (1996) and Girardi et al. (1997a). 
}
\end{figure*}

\end{document}